\documentclass[pre,preprint,showpacs]{revtex4}

\usepackage{epsfig}
\usepackage{graphics}
\usepackage{amssymb}

\begin{document}

\title{Drift by Dichotomous Markov Noise}

\date{\today}

\author{I. Bena}
\affiliation{D\'epartement de Physique Th\'eorique, Universit\'e de Gen\`eve,
CH-1211 Gen\`eve 4, Switzerland}

\author{C. Van den Broeck}
\affiliation{Limburgs Universitair Centrum, B-3590 Diepenbeek, Belgium}

\author{R. Kawai}
\affiliation{Department of Physics, University of Alabama at Birmingham,
Birmingham,  AL 35294}

\author{Katja Lindenberg}
\affiliation{Department of Chemistry and Biochemistry 0340 and Institute
for Nonlinear Science, University of California, San Diego,
La Jolla, CA 92093}

\begin{abstract}
We derive explicit results for the asymptotic probability density and drift
velocity in systems driven by dichotomous Markov noise, including the
situation in
which
the asymptotic dynamics crosses {\em unstable} fixed points. The results
are illustrated on
the problem of the rocking ratchet.
\end{abstract}

\pacs{05.40.-a,02.50.-r}
\maketitle


\section{Introduction}

Brownian motion is one of the paradigms of
statistical mechanics. Following the seminal works of
Einstein, Langevin and Smoluchowsky, a detailed
mathematical analysis can be made on the basis of  Langevin or
Fokker-Planck equations. While Brownian motion and
its ``time derivative", Gaussian white noise, are  stochastic processes
of fundamental importance,
the dichotomous Markov process
(see, e.g., \cite{kampen})
has its own virtues and interest. First -- and this is our central
contention here  --  systems driven by dichotomous noise can often be
described in full analytic
detail.
Second, dichotomous noise reduces to white shot noise and Gaussian
white noise in the appropriate limits \cite{vandenbroeck83}.
Third, it can  either mimic  the effects of finite correlation time of
the noise, or it may directly provide a good representation of an
actual physical situation (such as, for example, thermal transitions
between two configurations or states).
Finally, it has the advantage that it can easily be implemented as an
external noise with finite support.

Most of the  results for dynamics driven by
dichotomous Markov noise are limited to systems with a single
scalar variable $x(t)$.  
The rate of change of this variable switches
at random between the ``+" dynamics, $\dot{x}(t)=f_{+}(x)$, and
the ``--" dynamics,
$\dot{x}(t)=f_{-}(x)$ (the dot stands for the temporal derivative).
This dynamics can be described by the following stochastic differential
equation:
\begin{equation}
\dot{x}(t)=\displaystyle\frac{f_{+}(x)+f_{-}(x)}{2}+
\displaystyle\frac{f_{+}(x)-f_{-}(x)}{2}\xi(t),
\label{FOTP1}
\end{equation}
where $\xi(t)$ is a realization of the
dichotomous Markov process  taking the values $\pm 1$, with
transition rates between these values equal to
$k_{+}$ and $k_{-}$ respectively.
Several physical applications, including persistent random walks, nuclear
magnetic resonance,
chromatography,
Taylor dispersion, and kinetic theory,
have been discussed when the speeds $f_{+}$ and
$f_{-}$ are constants, and exact time-dependent results can be obtained
in this case \cite{linear}.
Somewhat surprisingly,
the full time-dependent solution for linear
$f_{+}(x)$ and $f_{-}(x)$  is not available, see \cite{nonlinear} for a
detailed discussion.
In the case of nonlinear dynamics,  steady state properties can be
calculated and have notably been studied in detail in the context of
noise induced transitions \cite{horsthemke} and  noise induced phase
transitions \cite{nit}. Specific dynamic properties have also been
obtained. A significant effort has gone into the calculation of 
first passage time moments \cite{MFPT}
and of  transition rates, cf.  Kramers' rate for thermal
escape {\cite{reimann96} and resonant activation \cite{ra}.
Furthermore, when $f_{+}(x)$ and
$f_{-}(x)$ are periodic another important dynamic property,
namely,
the asymptotic drift
velocity,
can be extracted from steady state results by working
with periodic boundary conditions. This technique is of particular
interest in the context
of Josephson junctions and  Brownian motors
\cite{magnasco,doering,gitterman97-98,stokesdrift}. However, contrary to
the claims made in some of these papers, the problem of the first
passage time moments, and the related issue of finding the asymptotic
drift velocity \cite{diffusion},
was not solved  in the most general case.  Indeed, with a few
exceptions \cite{behn,zapata,bala,czernik}, all
the results that
have been obtained 
exclude the cases when one or both of the ``$\pm$" dynamics
have {\em unstable} fixed points, and thus 
do not consider the possibility of crossing  unstable fixed points in the
long time dynamics.
The technical subtleties were first highlighted
and  discussed in detail in a recent paper
\cite{shortPRE}.  In view of the broad applicability and importance of
dichotomous
Markov noise, a comprehensive review of the results for
the asymptotic drift velocity in periodic systems driven by such a noise
is called for
and is the subject of this paper. As a first illustration
we also apply these results to the
calculation of the drift for the rocking ratchet.


\section{Solving the master equation}
\label{FOTP}

The  calculation of the asymptotic drift velocity in periodic systems is
most easily
carried out by starting from the
master equation, equivalent to (\ref{FOTP1}), for the probability densities
$P_{+}(x,t)$
and $P_{-}(x,t)$ to be
at $x$
at time $t$, if $\xi=+1$ and $-1$ respectively:
\begin{eqnarray}
&& \frac{\partial P_{+}(x,t)}{\partial
t}=-\frac{\partial}{\partial x}
\left[f_{+}(x)P_{+}(x,t)\right]-k_{+}P_{+}(x,t)+k_{-}P_{-}(x,\
,t),
\nonumber\\
&& \frac{\partial P_{-}(x,t)}{\partial t}=-\frac{\partial}{\partial x}
\left[f_{-}(x)P_{-}(x,t)\right]-k_{-}P_{-}(x,t)+k_{+}
P_{+}(x,t)
.
\label{FOTP3}
\end{eqnarray}
We assume that $f_{+}(x)$ and
$f_{-}(x)$ are  continuous functions of their argument and that
they are both periodic, i.e.,
\begin{equation}
f_{+}(x)=f_{+}(x+L)\,\,\,\, \mbox{and}\,\,\,\, f_{-}(x)=f_{-}(x+L),
\forall x.
\label{FOTP2}
\end{equation}
In order to extract the long-time average drift speed,
it is
sufficient to study the
steady state properties of  Eq. (\ref{FOTP3}) for
$x \in [0,L]$ with periodic boundary conditions.  To show this, we
introduce the steady state quantities
$P(x)=P_{+}(x)+P_{-}(x) $
(which represents the probability density for being at $x$
regardless the value of $\xi$)
\footnote{ We shall assume that $P(x)$ is unique, i.e., that the system is
ergodic. The necesary condition for this 
(that there are not several attractors
of the dynamics) will be clarified subsequently.}, and
$p(x)=k_{+} P_{+}(x)-k_{-}P_{-}(x) $.
From the summation of the two equations in (\ref{FOTP3})
one immediately concludes that for the asymptotic (steady) state the
probability flux $J$
associated with $P(x)$,
\begin{equation}
J=\displaystyle\frac{k_{+}f_{-}(x)+k_{-}f_{+}(x)}{k_{+}+k_{-}}
P(x)+\displaystyle\frac{f_{+}(x)-f_{-}(x)}{k_{+}+k_{-}}p(x),
\label{FOTP4}
\end{equation}
is a constant.  The asymptotic drift velocity is then simply given by:
\begin{eqnarray}
\langle \dot{x} \rangle=\int_0	^L
\left[f_{+}(x)P_+(x)+f_{-}(x)P_-(x)\right]dx=LJ.
\label{FOTP5}
\end{eqnarray}
By substracting the equations in (\ref{FOTP3}) (multiplied, respectively, by
$k_{+}$ and $k_{-}$),
one obtains, in the asymptotic steady state,
the following first-order differential equation  for $p(x)$:
\begin{eqnarray}
\frac{d}{dx}\left\{k_{+}k_{-}[f_{+}(x)-f_{-}(x)]P(x)+
\left[k_{+}f_{+}(x)+k_{-}f_{-}(x)\right]p(x)\right\}
+(k_{+}+k_{-})^2p(x)=0.
\label{FOTP6}
\end{eqnarray}
Equations (\ref{FOTP4}) and (\ref{FOTP6}) have to be solved by imposing
the conditions of {\it continuity} for $P(x)$ and $p(x)$
(or, at least, the condition
that they have no more than {\it integrable singularities}) on
$[0,L]$, {\it periodicity},
$P(x)=P(x+L)$ and $p(x)=p(x+L),\forall x$,
and {\it normalization} for $P(x)$,
$\displaystyle\int_0	^L	P(x)dx=1$.

All these elements allow the determination of the 
unique steady state solution
$P(x)$ , the stationary probability flux $J$, and hence the
corresponding dynamic quantity of interest to us, namely, the
asymptotic drift velocity $\langle\dot{x}\rangle$. 
As was pointed out
in detail in \cite{shortPRE}, the
situation is entirely different, both physically and mathematically,
when one or both of the ``$\pm$" dynamics have {\em unstable} fixed points,
depending on whether the system can cross (or not) these 
{\em unstable} fixed
points in the long time limit.
In fact, while this issue was also mentioned  in \cite{behn}
and \cite{zapata}, a full and detailed discussion 
was first given for a specific
example with multiplicative dichotomous noise in \cite{shortPRE}. In the
following we focus on the presentation and discussion of
the final results for the asymptotic probability density and drift
velocity, relegating the technical details to the Appendix.


\section{Main results}
\label{main}

Since we assume that
the functions $f_{\pm}(x)$ are continuous, and in view of the
periodicity,  fixed points in the separate dynamics
$\dot{x}=f_{+}(x)$ and
$\dot{x}=f_{-}(x)$ will always appear in pairs.
In the following we will present the final results for the three simplest
cases that can
occur, namely no fixed points, cf.
Section \ref{maina}, one of the dynamics has two fixed points and the
other none, Section \ref{mainb}, and both dynamics have two fixed points,
Section \ref{mainc}.


\subsection{No fixed points}
\label{maina}

The stationary probability
density $P(x)$ reads:
\begin{eqnarray}
P(x)&=&\displaystyle\frac{\langle\dot{x}\rangle\left[f_{+}(x)-f_{-}(x)\right
]}
{Lf_{+}(x)f_{-}(x)\left[\exp\left(\displaystyle\int_0^Ldz(k_{+}/f_{+}(z)+
\,k_{-}/f_{-}(z))\right)
-1\right]}\times\nonumber\\
&&\times\int_x^{x+L}dz
\left[\frac{k_{+}+k_{-}}{f_{+}(z)-f_{-}(z)}+
\left(\frac{f_{+}(z)+f_{-}(z)}{2[f_{+}(z)-f_{-}(z)]}\right)'\right]
\times\nonumber\\
&&\times\exp\left(-\int_z^x dw (k_{+}/f_{+}(w)+k_{-}/f_{-}(w))\right)
\label{simP}
\end{eqnarray}
and the mean asymptotic velocity is given by:
\begin{eqnarray}
\displaystyle{\langle\dot{x}\rangle} &=&
L\left[\exp\left(\displaystyle\int_0^Ldz(k_{+}/f_{+}(z)+k_{-}/f_{-}(z))\right)
-1\right]\left\{\int_0^Ldx\displaystyle
\frac{f_{+}(x)-f_{-}(x)}{f_{+}(x)f_{-}(x)}\right.\times\nonumber\\
&&\left.\times\int_x^{x+L}dz
\left[\frac{k_{+}+k_{-}}{f_{+}(z)-f_{-}(z)}+
\left(\frac{f_{+}(z)+f_{-}(z)}{2[f_{+}(z)-f_{-}(z)]}\right)^{'}\right]\times
\right.\nonumber\\
&&\left.\times\exp\left(-\int_z^x dw (k_{+}/f_{+}(w)+k_{-}/f_{-}(w))\right)
\right\}^{-1}.
\label{simvelocity}
\end{eqnarray}
Here $(...)^{'}$ stands for the derivative with respect to $x$.
Note that the above expressions for both $P(x)$ and $\langle\dot{x}\rangle$
display the required symmetry between the ``+" and ``--" dynamics.
The above expression reduces to the one given earlier in the literature for
the particular
case of additive dichotomous noise [$f_{+}(x)-f_{-}(x)$=constant, cf.
\cite{doering}].


\subsection{One of the dynamics has two fixed points in $[0,L)$}
\label{mainb}

We suppose that the ``+" dynamics has two fixed points
$x_1 < x_2$ in $[0,L)$, $x_1$ being {\em stable}
[$f_{+}(x_1)=0,\;
f_{+}'(x_1)<0$],
while $x_2$ is {\em unstable}
[$f_{+}(x_2)=0,\;f_{+}'(x_2)>0$].
The ``--" dynamics has no fixed
points.

As explained in detail in Appendix \ref{case1b} and in \cite{shortPRE}, there
exists {\em exactly one} solution $P(x)$ of the
Eqs. (\ref{FOTP4}) and
(\ref{FOTP6}) that is physically and mathematically acceptable, and for $x\in
[x_1,x_1+L]$ it is given by the following expression:
\begin{eqnarray}
P(x)&=&\displaystyle\frac{\langle\dot{x}\rangle}{L}
\left|\displaystyle\frac{f_{+}(x)-f_{-}(x)}{f_{+}(x)f_{-}(x)}\right|
\;\;\displaystyle\int_{x_2}^{x}dz\;\mbox{sgn}\left[
\displaystyle\frac{f_{+}(z)f_{-}(z)}{f_{+}(z)-f_{-}(z)}\right]
\times\nonumber\\
&&\times
\left[\displaystyle\frac{k_{+}+k_{-}}{f_{+}(z)-f_{-}(z)}+
\displaystyle\left(\frac{f_{+}(z)+f_{-}(z)}{2\left[f_{+}(z)-f_{-}(z)
\right]}\right)^{'}\right]\times\nonumber\\
&&\times\exp\left(-\int_z^x dw (k_{+}/f_{+}(w)+k_{-}/f_{-}(w))\right).
\label{twopointsP}
\end{eqnarray}
As discussed in Appendix \ref{case1b},
$P(x)$ is finite and continuous
throughout the interval $(x_1,x_1+L)$, with, in particular, at the
unstable
fixed point $x_2$
\begin{equation}
\lim_{x\nearrow x_2}P(x)= \lim_{x\searrow x_2}P(x)=
\displaystyle\frac{\langle\dot{x}\rangle}{L}\;
\displaystyle\frac{
(k_{+}+k_{-})/f'_{+}(x_2)+1}
{f_{-}(x_2)\left[k_{+}/f'_{+}(x_2)+1\right]}.
\label{twopointsx2}
\end{equation}
Note  that positivity of
$P(x)$ implies that  $f_{-}(x)$ and $\langle\dot{x}\rangle$ must have the
same sign. Hence
the direction of the asymptotic drift is the same as that of the ``--"
dynamics,
which is physically obvious.

At the stable fixed point $x_1$, $P(x)$ is either continuous when
$k_{+}/|f'_{+}(x_1)|>1$, i.e.,
\begin{equation}
\lim_{x\searrow x_1} P(x)= \lim_{x\nearrow (x_1+L)} P(x)=
\displaystyle\frac{\langle\dot{x}\rangle}{L}\;
\displaystyle
\frac{(k_{+}+k_{-})/|f'_{+}(x_1)|-1}
{f_{-}(x_1)\left[k_{+}/|f'_{+}(x_1)|-1\right]},
\label{case113}
\end{equation}
or divergent but integrable when $k_{+}/|f'_{+}(x_1)|\leqslant1$
[namely, $k_{+}/|f'_{+}(x_1)|<1$ corresponds to a power-law divergence,
while $k_{+}/|f'_{+}(x_1)|=1$ corresponds to a ``marginal" logarithmic-like
integrable divergence]. This result is consistent with the physical
intuition that probability density builds up near a stable
fixed point, especially
when the switching rate is low.

The normalization condition
$\displaystyle\int_{x_1}^{x_1+L} P(x)dx=1$, together with Eq.
(\ref{FOTP5}) lead
to the following  expression
for the mean velocity:
\begin{eqnarray}
\langle\dot{x}\rangle&=&L\left\{
\int_{x_1}^{x_1+L}dx\displaystyle
\left|\displaystyle\frac{f_{+}(x)-f_{-}(x)}{f_{+}(x)f_{-}(x)}\right|
\;\;\displaystyle\int_{x_2}^{x}dz\;\mbox{sgn}\left[
\displaystyle\frac{f_{+}(z)f_{-}(z)}{f_{+}(z)-f_{-}(z)}\right]
\right.\times\nonumber\\
&&\times
\left[\displaystyle\frac{k_{+}+k_{-}}{f_{+}(z)-f_{-}(z)}+
\displaystyle\left(\frac{f_{+}(z)+f_{-}(z)}{2\left[f_{+}(z)-f_{-}(z)
\right]}\right)^{'}\right]\times\nonumber\\
&&\left.\times\exp\left(-\int_z^x dw
(k_{+}/f_{+}(w)+k_{-}/f_{-}(w))\right)
\right\}^{-1}.
\label{case116}
\end{eqnarray}
This expression reduces to the one mentioned in \cite{zapata} for the particular
case of a symmetric additive dichotomous noise.


\subsection{Each of the alternating dynamics has two fixed points in $[0,L)$}
\label{mainc}

Consider as above $x_1 < x_2$ as the stable,
respectively unstable fixed points of the ``+" dynamics, and suppose now that
the ``--"
dynamics also has two fixed points, $x_3<x_4$.
Depending on the relative positions of these four
points, as well as on their nature (stable or unstable), two different
types of situations
might occur.

(i) In all the situations in which $x_1$ has as direct
neighbor another {\em stable} fixed point,  it is physically clear that
the asymptotic dynamics settles  into a random alternating motion between
these
points,
so that they delimit the interval in which the steady-state probability
density is
non-zero \cite{horsthemke}. Obviously in this case
$\langle\dot{x}\rangle=0$, i.e.,
there is no net flux of the particles, while $P(x)$ is given by
\begin{equation}
P(x)=C\;\;\displaystyle\frac{f_{+}(x)-f_{-}(x)}{f_{+}(x)f_{-}(x)}
\;\exp\left(-\int_{x_0}^x dz (k_{+}/f_{+}(z)+k_{-}/f_{-}(z))\right),
\label{case118}
\end{equation}
for $ x$ and $x_0$  lying in between the two stable fixed
points, while the constant $C$ is determined through the normalization
condition
for $P(x)$.

(ii) The cases that correspond to
an alternation
of the stable and unstable
fixed points are of more interest to us, as they lead to a nontrivial
behavior of
$P(x)$ and to a nonzero flow of the particles. Without loss of generality
we suppose that
$0<x_1$ (s) $<x_3$ (u) $<x_4$ (s) $<x_2$ (u) $<L$,
with the following results:
\begin{equation}P(x)=\left\{\begin{array}{l}
\displaystyle\frac{\langle\dot{x}\rangle}{L}\;
\left|\displaystyle\frac{f_{+}(x)-f_{-}(x)}{f_{+}(x)f_{-}(x)}\right|
\;\;\displaystyle\int_{x_3}^{x}dz\;\mbox{sgn}\left[
\displaystyle\frac{f_{+}(z)f_{-}(z)}{f_{+}(z)-f_{-}(z)}\right]
\times\\
\hspace{1cm}\times
\left[\displaystyle\frac{k_{+}+k_{-}}{f_{+}(z)-f_{-}(z)}+
\displaystyle\left(\frac{f_{+}(z)+f_{-}(z)}{2\left[f_{+}(z)-f_{-}(z)
\right]}\right)^{'}\right]\times\\
\hspace{1cm}\times\exp\left(-\displaystyle\int_z^x dw
(k_{+}/f_{+}(w)+k_{-}/f_{-}(w))\right)\\
\hspace{7cm}\mbox{for}\,\,\,\, x\in(x_1,x_4) \\
\\
\displaystyle\frac{\langle\dot{x}\rangle}{L}
\left|\displaystyle\frac{f_{+}(x)-f_{-}(x)}{f_{+}(x)f_{-}(x)}\right|
\;\;\displaystyle\int_{x_2}^{x}dz\;\mbox{sgn}\left[
\displaystyle\frac{f_{+}(z)f_{-}(z)}{f_{+}(z)-f_{-}(z)}\right]
\times\\
\hspace{1cm}\times
\left[\displaystyle\frac{k_{+}+k_{-}}{f_{+}(z)-f_{-}(z)}+
\displaystyle\left(\frac{f_{+}(z)+f_{-}(z)}{2\left[f_{+}(z)-f_{-}(z)
\right]}\right)^{'}\right]\times\\
\hspace{1cm}\times\exp\left(-\displaystyle\int_z^x dw
(k_{+}/f_{+}(w)+k_{-}/f_{-}(w))\right)\\
\hspace{7cm}\mbox{for}\,\,\,\, x\in(x_4,x_1+L)
\end{array}
\right..
\label{case120}
\end{equation}
$P(x)$ is continuous throughout $(x_1,x_4)$ and $(x_4,x_1+L)$
and, in particular, at the unstable fixed points $x_2$ and $x_3$ it
takes, respectively,  the values  (\ref{twopointsx2}) and
\begin{equation}
\lim_{x\nearrow x_3} P(x)= \lim_{x\searrow x_3} P(x)=\displaystyle
\displaystyle\frac{\langle\dot{x}\rangle}{L}\;
\frac{(k_{+}+k_{-})/f'_{-}(x_3)+1}
{f_{+}(x_3)\left[k_{-}/f'_{-}(x_3)+1\right]}.
\end{equation}

The periodicity of $P(x)$ is connected with the behavior at the stable fixed
point $x_1$ that was already discussed in the previous section,
see Eq.
(\ref{case113}).

Concerning the behavior of $P(x)$ at the other stable fixed point, $x_4$,
it is either divergent
(but integrable) for $k_{-}/|f'_{-}(x_4)|\leqslant1$,
or it is continuous,
\begin{equation}
\lim_{x\searrow x_4} P(x)= \lim_{x\nearrow x_4} P(x)=
\displaystyle\frac{\langle\dot{x}\rangle}{L}\;
\displaystyle
\frac{(k_{+}+k_{-})/|f'_{-}(x_4)|-1}
{f_{+}(x_4)\left[k_{-}/|f'_{-}(x_4)|-1\right]},
\end{equation}
when $k_{-}/|f'_{-}(x_4)|>1$.

Imposing the normalization condition for $P(x)$, one obtains the corresponding
asymptotic drift velocity:
\begin{eqnarray}
\langle\dot{x}\rangle&=&L\left\{\int_{x_1}^{x_4}dx
\left|\displaystyle\frac{f_{+}(x)-f_{-}(x)}{f_{+}(x)f_{-}(x)}\right|
\;\;\displaystyle\int_{x_3}^{x}dz\;\mbox{sgn}\left[
\displaystyle\frac{f_{+}(z)f_{-}(z)}{f_{+}(z)-f_{-}(z)}\right]\right.
\times\nonumber\\
&&\hspace{1cm}\times
\left[\displaystyle\frac{k_{+}+k_{-}}{f_{+}(z)-f_{-}(z)}+
\displaystyle\left(\frac{f_{+}(z)+f_{-}(z)}{2\left[f_{+}(z)-f_{-}(z)
\right]}\right)^{'}\right]\times\nonumber\\
&&\hspace{1cm}\times\exp\left(-\int_z^x dw
(k_{+}/f_{+}(w)+k_{-}/f_{-}(w))\right)+\nonumber\\
&&+\int_{x_4}^{x_1+L}dx
\left|\displaystyle\frac{f_{+}(x)-f_{-}(x)}{f_{+}(x)f_{-}(x)}\right|
\;\;\displaystyle\int_{x_2}^{x}dz\;\mbox{sgn}\left[
\displaystyle\frac{f_{+}(z)f_{-}(z)}{f_{+}(z)-f_{-}(z)}\right]
\times\nonumber\\
&&\hspace{1cm}\times
\left[\displaystyle\frac{k_{+}+k_{-}}{f_{+}(z)-f_{-}(z)}+
\displaystyle\left(\frac{f_{+}(z)+f_{-}(z)}{2\left[f_{+}(z)-f_{-}(z)
\right]}\right)^{'}\right]\times\nonumber\\
&&\hspace{1cm}\left.\times\exp\left(-\int_z^x dw
(k_{+}/f_{+}(w)+k_{-}/f_{-}(w))\right)
\right\}^{-1},
\label{case124}
\end{eqnarray}
which completes the discussion.

All these results are suitable for a rapid generalization to cases when
both dynamics have several pairs of fixed points. In particular, note that 
when there are several pairs of {\em adjacent stable fixed points},
the system is no longer ergodic. Depending on the initial conditions, 
the asymptotic motion of the particles is limited to one or another basin of
attraction. These attractors are represented, on the real axis, by the intervals
between such pairs of adjacent stable fixed points. We shall not go into further
details here.

\section{Rocking ratchet}
The general formulas that were derived above still involve triple
integrals. One interesting
case for which explicit results can be obtained is that of
piece-wise constant  functions $f_{+}$ and
$f_{-}$. To illustrate the results for this case,  we will focus on one of
the   paradigms
for   Brownian motors, namely the rocking ratchet
\cite{rr}. 
\begin{figure}
\includegraphics{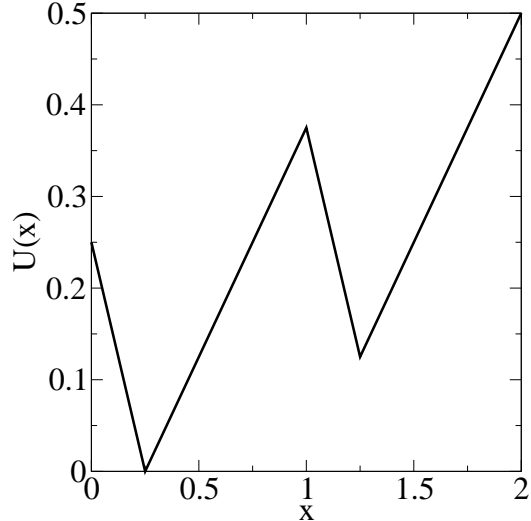}
\caption{The potential U(x) used for numerical calculations on the rocking
ratchet
($f_1$=1.0, $f_2$=0.5, $L$=1.0 and $L_1$=0.25).}
\label{fig:ux}
\end{figure}
In this problem an overdamped particle is gliding in a
deterministic
sawtooth-like potential
$U(x)$ subject to an additional dichotomous forcing that
switches between the
values $A_{+}$ and $-A_{-}$. The equation of motion is of the form:
\begin{equation}
\dot{x}=f(x)+\displaystyle\frac{A_{+}-A_{-}}{2}+
\displaystyle\frac{A_{+}+A_{-}}{2}\xi(t)\,,
\label{langevin-eq}
\end{equation}
with $f(x)=-U'(x)$,
and  thus $f_{\pm}(x)=f(x)\pm A_{\pm}$.
Without loss of generality, we assume that the mean value of the forcing  is
zero, i.e., $A_{+}/k_{+}=A_{-}/k_{-}$, and, in order to fix ideas, we suppose that
$A_{-}\geq A_{+}>0$ (the equality corresponds to a symmetric
dichotomous noise).
$f(x)$ has a periodic ``block-wave" profile:
\begin{equation}
f(x)=
\left\{
\begin{array}{rl}
 f_{1}& \,\,\,\, \textrm{for}\quad  x \in [0,L_{1})\\
-f_{2}& \,\,\,\, \textrm{for}\quad  x \in [L_{1},L_{1}+L_{2}),
\end{array}
\right.
\label{bw}
\end{equation}
with $f(x+L)=f(x)$, where
$L=L_{1}+L_{2}$; also,
$f_{1}$ and $f_{2}$  are supposed to be two positive constants,  with,
e.g.,  $f_{2}<f_{1}$.
Although analytical calculation are done in full generality,
all our numerical calculations were made for the potential represented in Fig.
\ref{fig:ux}, i.e., we considered the case of a negative bias. In this case, the
mean asymptotic velocity is determined both by a ratchet-like effect (if
present) and by the bias of the potential. Of course, one can also consider (as
usually done when studying ratchet effect) the case of an untilted potential,
that corresponds to the condition $f_{1} L_{1}=f_{2} L_{2}$.

We now proceed to a presentation of the results as they follow from the
more general discussion
presented earlier. Note that in the latter discussion $f_{\pm}(x)$ 
were assumed to be continuous.
Therefore, in order to be able to use the 
results obtained previously, we shall
first consider, for computational purposes, 
the following continuous, piece-wise
linear force profile: 
\begin{equation}
f(x)=
\left\{
\begin{array}{ll}
 f_{1}& \,\,\,\, \textrm{for}\quad  x \in [0,L_{1}-l)\\
 f_1-(f_1+f_2)(x-L_1+l)/l& \,\,\,\, \textrm{for}\quad  x \in [L_{1}-l,L_1)\\
-f_{2}& \,\,\,\, \textrm{for}\quad  x \in [L_{1},L_{1}+L_{2}-l)\\
 -f_2+(f_1+f_2)(x-L_1-L_2+l)/l& \,\,\,\, \textrm{for}\quad  x \in 
 [L_{1}+L_2-l,L_1+L_2).
\end{array}
\right.
\end{equation}
Afterwards we shall take the limit $l\rightarrow 0$ [corresponding to the 
``block-wave" profile (\ref{bw})] 
in the final expressions for the asymptotic probability 
density and the mean velocity.

\subsubsection{Strong forcing: no fixed points}
When $A_{-}>f_{1}$ and $A_{+}>f_{2}$,
running solutions appear for both tilts $+A_{+}$ and $-A_{-}$, and
there is no fixed
point in any of the separate dynamics ``+'' and ``-''. Applying the results
obtained in
Section  \ref{maina}, one finds the following expression for the
probability density in the two subintervals of one period $L$:
\begin{equation}
P(x)=
\left\{
\begin{array}{l}
\displaystyle\frac{\langle\dot{x}\rangle}{L}
\left\{
     \displaystyle\frac{-\, (f_{1}+f_{2}) A_{+} A_{-}(e^{\phi_2}-1)}
     {f_{1} f_{2} (A_{+}+f_{1}) (A_{-}-f_{1})(e^{\phi_2-\phi_1}-1)} 
     \exp(-\phi_1 |x-L_{1}|/L_{1})+
     \frac{1}{f_{1}}
\right\},\\
\hfill\textrm{for} \quad x \in [0,L_{1})\\
\\
\displaystyle\frac{\langle\dot{x}\rangle}{L}\;
\left\{
     \displaystyle\frac{(f_{1}+f_{2}) A_{+} A_{-}(e^{\phi_1}-1)}
     {f_{1} f_{2} (A_{+}-f_{2}) (A_{-}+f_{2})(e^{\phi_1-\phi_2}-1)} 
     \exp[-\phi_2 (x-L_{1})/L_{2}]-
     \frac{1}{f_{2}}
\right\},\\
\hfill\textrm{for} \quad x \in [L_{1},L)
\end{array}
\right. .
\label{eq:px1}
\end{equation}
We used here the dimensionless quantities
\begin{equation}
\phi_1=\displaystyle\frac{(k_{+}+k_{-})f_{1}L_{1}}
{(A_{-}-f_{1})(A_{+}+f_{1})}
\,\,\,\,\mbox{and}\,\,\,\,
\phi_2=\displaystyle\frac{(k_{+}+k_{-})f_{2}L_{2}}
{(A_{-}+f_{2})(A_{+}-f_{2})}.
\end{equation}
Because of the first-order discontinuity of $f(x)$ at
$x=L_{1},L$, $P(x)$
is also discontinuous at these points.

The mean asymptotic velocity reads:
\begin{equation}
\displaystyle{\langle\dot{x}\rangle}=L
\left[
   \left(
      \displaystyle\frac{L_{1}}{f_{1}}-\frac{L_{2}}{f_{2}}
   \right)+
   \displaystyle\frac{A_{+}A_{-}}{k_{+}+k_{-}}
   \left(
      \displaystyle\frac{f_{1}+f_{2}}{f_{1}f_{2}}
   \right)^2
   \left(
      \displaystyle\frac{1-e^{-\phi_1}-e^{\phi_2}+e^{\phi_2-\phi_1}}
                        {e^{\phi_2-\phi_1}-1}
   \right)
   \right]^{-1},
   \label{eq:v1}
\end{equation}
and it is determined by the interplay between 
the characteristics of
the noise and those of the potential (in particular, its bias, if any).
Some limit cases of interest include:

(i) When $k_{\pm}\rightarrow\infty$ (but $A_{\pm}$ finite) one finds,
of course,
$\langle\dot{x}\rangle=0$ -- the noiseless (deterministic) result.

(ii) The quenched noise-limit $k_{\pm}\rightarrow0$
[i.e., a fraction $A_{-}/(A_{+}+A_{-})$ of the particles,
chosen at random, are subjected to a constant external forcing $A_{+}$, while
the remaining
ones  are subjected to an external forcing $-A_{-}$] results in the
following mean
velocity:
\begin{equation}
{\langle\dot{x}\rangle}=L\;
\displaystyle
\frac{f_{1}f_{2}(L_{1}f_{2}-L_{2}f_{1})-A_{+}A_{-}(L_{1}f_{1}-L_{2}f_{2})-
Lf_{1}f_{2}(A_{+}-A_{-})}
{(L_{1}f_{2}-L_{2}f_{1})^2-L(A_{+}-A_{-})(L_{1}f_{2}-L_{2}f_{1})-
L^{2}A_{+}A_{-}}.
\end{equation}

(iii) One obtains  {\em white shot noise} \cite{vandenbroeck83} by taking
the limit
$A_{-}\rightarrow\infty$, $k_{-}\rightarrow\infty$,
such that
$A_{-}/k_{-}=A_{+}/k_{+}\equiv\lambda\neq0$ ($A_{+}$ and $k_{+}$ being
finite).
The mean
velocity is then
\begin{equation}
\langle\dot{x}\rangle=L\left[\left(\displaystyle\frac{L_{1}}{f_{1}}-
\displaystyle\frac{L_{2}}{f_{2}}\right)+\lambda A_{+}
\left(\displaystyle\frac{f_{1}+f_{2}}{f_{1}f_{2}}\right)^2\;
\displaystyle\frac{1-e^{-\phi_1}-e^{\phi_2}+e^{\phi_2-\phi_1}}
{e^{\phi_2-\phi_1}-1}\right]^{-1},
\label{shot1}
\end{equation}
where $\phi_1=f_{1}L_{1}/[\lambda(A_{+}+f_{1})]$ and
$\phi_2=f_{2}L_{2}/[\lambda(A_{+}-f_{2})]$. In the case of a symmetric
potential
($f_{1}=f_{2}$ and $L_{1}=L_{2}$), one recovers the result in \cite{shot}.

Moreover, 
for very large values of $A_+$ (and finite $\lambda$) one 
recovers the noiseless limit,
namely,
$\langle\dot{x}\rangle \approx (f_1\,L_1-f_2\,L_2)/L$
(that is strictly due to the bias of the potential).}

(iv) Finally, the white noise limit:
$A_{+}=A_{-}\equiv A\rightarrow\infty$
and $k_{+}=k_{-}\equiv k\rightarrow\infty$, with $A^2/2k=D$
finite.
Then
\begin{equation}
\phi_1=\displaystyle\frac{f_{1}L_{1}}{D}\,\,\,\, \textrm{and}\,\,\,\,
\phi_2=\displaystyle\frac{f_{2}L_{2}}{D},
\end{equation}
and the mean velocity (\ref{eq:v1}) takes the well-known form 
(see, e.g. \cite{risken}):
\begin{equation}
\langle\dot{x}\rangle=L
\left[\left(\displaystyle\frac{L_{1}}{f_{1}}-\frac{L_{2}}{f_{2}}\right)+D
\left(\displaystyle\frac{f_{1}+f_{2}}{f_{1}f_{2}}\right)^2
\left(\displaystyle\frac{1-e^{-\phi_1}-e^{\phi_2}+e^{\phi_2-\phi_1}}
{e^{\phi_2-\phi_1}-1}\right)\right]^{-1}.
\end{equation}
In the case of an unbiased potential one has now
$\phi_1=\phi_2=\phi$, implying a zero mean velocity 
(as required by the second law of thermodynamics)
and a probability
density profile
that assumes the Boltzmann form:
\begin{equation}
P(x)=
\displaystyle\frac{f_{1}f_{2}}{(f_{1}+f_{2})D(1-e^{-\phi})} \times
\left\{
\begin{array}{ll}
\exp(-\phi|x-L_{1}|/L_{1}),&\textrm{for}\quad x \in [0,L_{1}),\\
\\
\exp(-\phi(x-L_{1})/L_{2}),&\textrm{for}\quad x \in [L_{1},L).
\end{array}
\right.
\end{equation}

\subsubsection{Intermediate forcing: two fixed points}

For intermediate forcing, there are two situations
that might occur.

(1). When $A_{-}>f_{1}$ but $A_{+}<f_{2}$, the ``+" dynamics has two fixed
points, namely,
$x=L_{1}$ (stable, which corresponds to an asymmetric $\delta$ peak in
the probability density),
and $x=L$ (unstable, corresponds to a finite discontinuity in the
probability density).
One obtains from the results in Section  \ref{mainb}:
\begin{eqnarray}
P(x) = 
&-&\displaystyle\frac{\langle\dot{x}\rangle}{L}
\displaystyle
\frac{2A_{-}(A_{+}+f_{1})(e^{\phi_1}-1)}{k_{+}f_{1}(A_{+}+A_{-})}
\delta_{-}(L_{1}-x) \nonumber\\
&-&\displaystyle\frac{\langle\dot{x}\rangle}{L}
\displaystyle
\frac{2A_{-}(A_{+}-f_{2})(e^{\phi_2}-1)}{k_{+}f_{2}(A_{+}+A_{-})}
\delta_{+}(x-L_{1})
\nonumber\\
&&\nonumber\\
&-&
\left\{
\begin{array}{ll}
\displaystyle\frac{\langle\dot{x}\rangle}{L f_{1}}
\left[\displaystyle\frac{A_{-}e^{\phi_1}}{A_{-}-f_{1}}
\exp[-\phi_1 |x-L_{1}|/L_{1}]-1\right]
&\textrm{for}\quad x \in [0,L_{1})\\
\\
\displaystyle\frac{\langle\dot{x}\rangle}{L f_{2}}
\left[\displaystyle\frac{-A_{-}e^{\phi_2}}{A_{-}+f_{2}}
\exp[-\phi_2 (x-L_{1})/L_{2}]+1\right]
&\textrm{for}\quad x \in [L_{1},L).
\end{array}
\right.
\label{eq:px2}
\end{eqnarray}
Here $\displaystyle\delta_{\pm}(x)$ are the half
Dirac-delta functions
\footnote {This means
$\displaystyle\int_{-\infty}^{-0}\delta_{-}(x)f(x)dx=1/2
\lim_{x\nearrow 0} f(x)=1/2f(-0)$,
respectively
$\displaystyle
\int_{+0}^{\infty}\delta_{+}(x)f(x)dx=1/2
\lim_{x\searrow 0} f(x)=1/2f(+0)$,
and $\delta(x)=\delta_{-}(x)+\delta_{+}(x)$}.
A simple intuitive explanation for their appearance is related to the fact
that in the limit of the ``block-wave" force $f(x)$, Eq.~(\ref{bw}), a particle
can reach the stable fixed point $x=L_1$ (coming either from its left or from
its right) in a {\em finite time}, and, once there, it will stay at this
position for the reminder of the ``+" dynamics. 
Thus, in time average, this will
give rise to a finite weight at this precise point.
The corresponding mean velocity is then:
\begin{eqnarray}
{\langle\dot{x}\rangle}= &-& L
\left\{ 
\left(\displaystyle\frac{L_{2}}{f_{2}}-\frac{L_{1}}{f_{1}}\right)\right.
\nonumber\\
&+&\left.
\displaystyle\frac{A_{-}(A_{+}+f_{1})(e^{\phi_1}-1)}
{f_{1}}\left[\displaystyle\frac{1}{k_{+}(A_{+}+A_{-})}+
\displaystyle\frac{1}{(k_{+}+k_{-})f_{1}}\right]\right.\nonumber\\
&+&\left.
\displaystyle\frac{A_{-}(A_{+}-f_{2})(e^{\phi_2}-1)}
{f_{2}}\left[\displaystyle\frac{1}{k_{+}(A_{+}+A_{-})}-
\displaystyle\frac{1}{(k_{+}+k_{-})f_{2}}\right]\right\}^{-1}.
\label{eq:v2}
\end{eqnarray}

The direction of the mean velocity is, as expected, the same as
that of the ``--"
dynamics, i.e., $\langle\dot{x}\rangle<0$ in this case.

\begin{figure}
\centerline{\epsfxsize=10cm\epsfbox{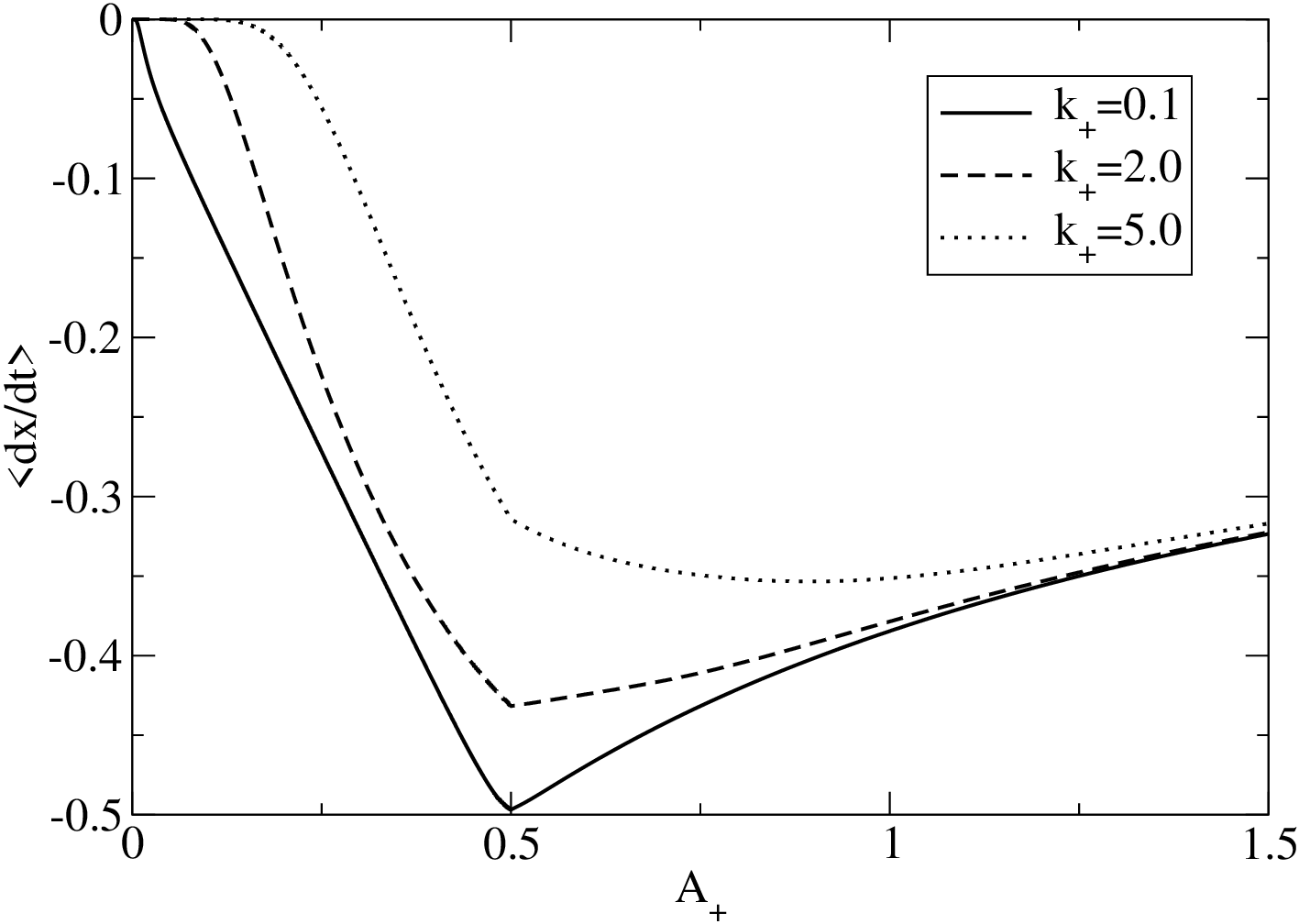}}
\caption{For the white shot noise limit [eqs. (\ref{shot1}), 
respectively (\ref{shot2})]:
The mean asymptotic velocity as a function of $A_+$ for various values
of the transition rate $k_+$. The sawtooth potential is the one 
represented in Fig. \ref{fig:ux}.}
\label{shot}
\end{figure}

One interesting limit in this case is
that of white shot noise:
$A_-\rightarrow \infty$, $k_-\rightarrow\infty$ so that $A_-/k_-=A_+/k_+\equiv
\lambda$ finite.  Our results are equivalent to those presented 
in \cite{czernik} for the probability density 
(although written under a closed,
compact form).
One obtains
the following expressions for the asymptotic velocity :
\begin{eqnarray}
\langle\dot{x}\rangle&=&-L\,\left[\left(\displaystyle\frac{L_2}{f_2}-
\displaystyle\frac{L_1}{f_1}\right)+
\displaystyle\frac{\lambda (A_++f_1)^2\,(e^{\phi_1}-1)}{A_+\,f_1^2}\right.
\left.-\displaystyle\frac{\lambda (A_+-f_2)^2\,(e^{\phi_2}-1)}{A_+\,f_2^2}
\right]^{-1}\;.
\label{shot2}
\end{eqnarray}
Figure \ref{shot} shows the variation of the average asymptotic velocity as a
function of $A_+$ [according to
Eq. (\ref{shot1}),
respectively (\ref{shot2})]
for various fixed values of the transition rate $k_+$. Note the peak and the
discontinuity in the slope of the
velocity at $A_+=f_2$,
connected with the
appearance of the fixed points when $A_+$ decreaseas below $f_2$. 

(2). When $A_{+}>f_{2}$ but $A_{-}<f_{1}$,  there are no fixed points in
the ``+" dynamics, but two fixed points in the ``--" dynamics, namely
$x=L_{1}$ (stable)
and
$x=L$ (unstable); together with the discontinuous character of $f(x)$
at these points, they lead, respectively, to an asymmetric
$\delta$-peak of $P(x)$ at $x=L_{1}$
and to a first-order discontinuity of $P(x)$ at $x=L$:
\begin{eqnarray}
P(x)&=&
\displaystyle\frac{\langle\dot{x}\rangle}{L}
\displaystyle
\frac{2A_{+}(A_{-}-f_{1})(e^{\phi_1}-1)}{k_{-}f_{1}(A_{+}+A_{-})}
\delta_{-}(L_{1}-x) \nonumber\\
&+&\displaystyle\frac{\langle\dot{x}\rangle}{L}
\displaystyle
\frac{2A_{+}(A_{-}+f_{2})(e^{\phi_2}-1)}{k_{-}f_{2}(A_{+}+A_{-})}
\delta_{+}(x-L_{1}) \nonumber\\
&&\nonumber\\
&+&
\left\{
\begin{array}{ll}
\displaystyle\frac{\langle\dot{x}\rangle}{Lf_{1}}
\left[\displaystyle\frac{-A_{+}e^{\phi_1}}{A_{+}+f_{1}}
\exp[-\phi_1 |x-L_{1}|/L_{1}]+1\right],
& \textrm{for} \quad x \in [0,L_{1})\\
\\
\displaystyle\frac{\langle\dot{x}\rangle}{Lf_{2}}
\left[\displaystyle\frac{A_{+}e^{\phi_2}}{A_{+}-f_{2}}
\exp[-\phi_2 (x-L_{1})/L_{2}]-1\right],
& \textrm{for} \quad x \in [L_{1},L).
\end{array}
\right.
\end{eqnarray}
This leads to a  mean velocity that reads:
\begin{eqnarray}
{\langle\dot{x}\rangle} &=& L\left\{
\left(\displaystyle\frac{L_{1}}{f_{1}}-\frac{L_{2}}{f_{2}}\right)
\right. \nonumber\\
&+&\left.
\displaystyle\frac{A_{+}(A_{-}-f_{1})(e^{\phi_1}-1)}{f_{1}}
\left[\displaystyle\frac{1}{k_{-}(A_{+}+A_{-})}-
\displaystyle\frac{1}{(k_{+}+k_{-})f_{1}}\right]\right.\nonumber\\
&+&\left.
\displaystyle\frac{A_{+}(A_{-}+f_{2})(e^{\phi_2}-1)}{f_{2}}
\left[\displaystyle\frac{1}{k_{-}(A_{+}+A_{-})}+
\displaystyle\frac{1}{(k_{+}+k_{-})f_{2}}\right]\right\}^{-1}.
\end{eqnarray}
Note that in this case $\langle\dot{x}\rangle>0$, 
as dictated by
the sign of the ``+" dynamics.
One can thus clearly realize the role of the 
ratchet effect if, e.g.,  one considers a potential with negative bias 
(like the one in Fig.
\ref{fig:ux}): indeed, in this case 
the mean velocity is directed against the
bias.
The above expression simplifies for the case of a symmetric dichotomic
noise $A_{+}=A_{-}\equiv A$ and $k_{+}=k_{-}\equiv k$, and reads:
\begin{equation}
\langle\dot{x}\rangle=L
\left[\left(\displaystyle\frac{L_{1}}{f_{1}}-\frac{L_{2}}{f_{2}}\right)-
\displaystyle\frac{(A-f_{1})^2(e^{\phi_1}-1)}{2kf_{1}^2}+
\displaystyle\frac{(A+f_{2})^2(e^{\phi_2}-1)}{2kf_{2}^2}\right]^{-1}.
\end{equation}

\subsubsection{Weak forcing}

Finally, when $A_{-}<f_{1}$ and $A_{+}<f_{2}$,  there are no freely running
solution in either of the dynamics. The points $x=L_{1},L$
are fixed points for both ``+" and ``--" dynamics.
One concludes  that the mean asymptotic velocity is zero, and all the
particles are
concentrated at the stable fixed point
$x=L_{1}$, i.e., $P(x)=\delta(x-L_{1})$. In this case additional
thermal
noise is needed to generate rectified motion. This problem has been solved for
adiabatically slow forcing \cite{magnasco,doering}.

\begin{figure}
\includegraphics{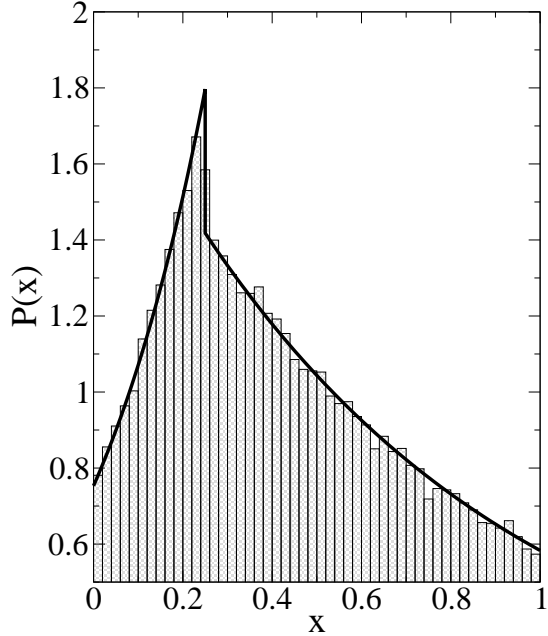}
\caption{Probability distribution $P(x)$ for the strong forcing case 
($A_+$=$A_-$=2.0, and
$k_+$=$k_-$=5.0; see Fig. \ref{fig:ux} for the parameters of $U(x)$.).
The solid line shows the
theoretical result (\ref{eq:px1}) and the shaded
histogram presents the results of numerical simulations.}
\label{fig:px1}
\end{figure}

\begin{figure}
\includegraphics{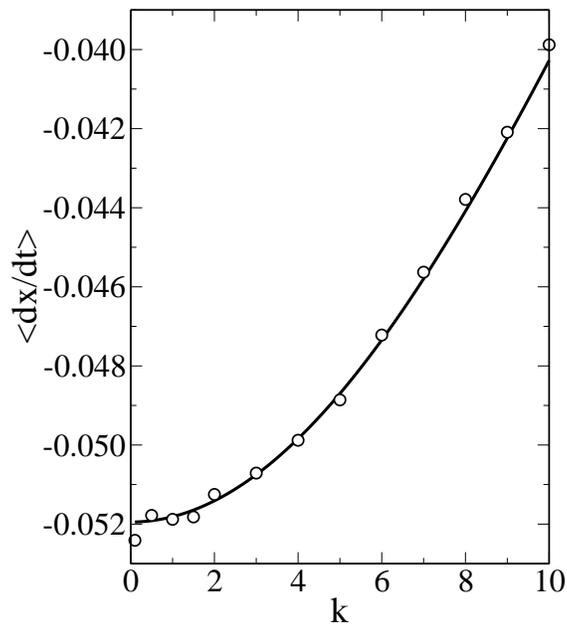}
\caption{Mean asymptotic velocity as a function of the transition rate $k$ for
the strong forcing
case (see Figs. \ref{fig:ux} and \ref{fig:px1} for parameter values). The solid line shows the
theoretical result (\ref{eq:v1}) and the open circles the results
of numerical simulations.}
\label{fig:v1}
\end{figure}

\subsubsection{Comparison with numerical simulations}

\begin{figure}
\includegraphics{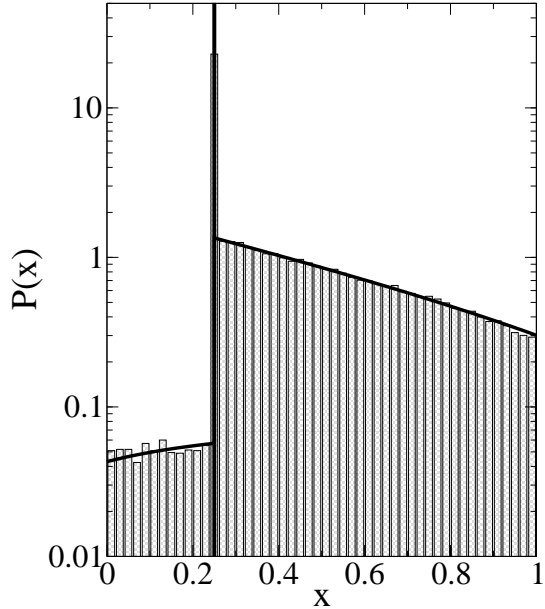}
\caption{Probability distribution $P(x)$ for the intermediate forcing case
($A_+$=$A_-$=0.75,
$k_+$=$k_-$=0.5; see Fig. \ref{fig:ux} for the parameters of $U(x)$).
The solid line shows the theoretical result (\ref{eq:px2}) and the shaded
histogram presents the results of numerical simulations.}
\label{fig:px2}
\end{figure}

\begin{figure}
\includegraphics{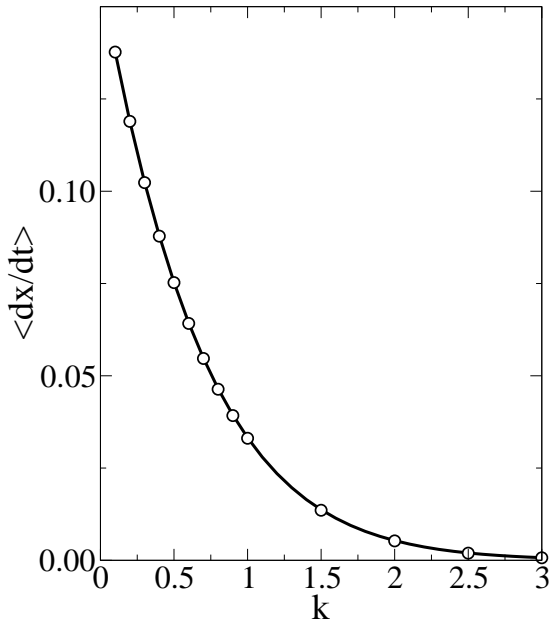}
\caption{Particle velocity as a function of the transition rate $k$ for
the intermediate forcing case
(see. Figs. \ref{fig:ux} and \ref{fig:px1} for parameter values). The solid line is the
theoretical result (\ref{eq:v2}) and the open circles result from
numerical simulations.}
\label{fig:v2}
\end{figure}

The Langevin equation
(\ref{langevin-eq}) is simulated numerically 
using 100,000 particles and the probability distribution is averaged over 10
snapshots. The biased potential shown in Fig. \ref{fig:ux} is used for all
numerical calculations.  
Figures \ref{fig:px1} and \ref{fig:v1} show, respectively,
the probability distribution $P(x)$ and the
average velocity $\langle\dot{x}\rangle$ for the strong forcing
case, and Figs. \ref{fig:px2} and \ref{fig:v2} for the intermediate
forcing case.  Agreement between theory and simulations is very good.
Discontinuities in $P(x)$ at $x=L_1$ and $x=L$ are clearly seen in
both cases.  Figure \ref{fig:px2} also illustrates the existence of a
delta peak at $x=L_1$. Note the phenomenon of {\it current reversal} when one
decreases the amplitude of the perturbation (for a fixed $k$), i.e., when one
passes from strong forcing (no fixed points, 
 cf. Fig. \ref{fig:v1}) 
to intermediate forcing (fixed
points appearing, 
cf. Fig. \ref{fig:v2}).

\section{Perspectives}

The results obtained above 
reinforce the impression that dichotomous noise can be put
on a par
with Gaussian white noise
as far as obtaining analytic results is concerned.
Along this line of thought,
we expect that one
can  obtain exact results for the first passage time moments when
unstable fixed points are crossed. Also one can extend the calculations for the
asymptotic  diffusion coefficient in periodic system subject to additive
Gaussian white noise
\cite{reimann01,lutz01}
to the case of dichotomous forcing via its
relationship to the
first two  moments of the first passage time \cite{diffusion}.
Furthermore, the results presented here can be directly applied
to various other problems,
including  Stokes' drift \cite{stokesdrift} and hypersensitive
transport
\cite{ht}.\\

\section{Appendix}

From the point of view of the mathematical steps involved in solving
Eqs.  (\ref{FOTP4}) and (\ref{FOTP6}), there are several
cases that have to be
considered separately:\\
{\bf Case I}, when the coefficient of $p(x)$
in Eq.  (\ref{FOTP4}), namely $[f_{+}(x)-f_{-}(x)]$,
has no zeroes in $[0,L)$,  one can solve this equation 
for $p(x)$;
\\ {\bf Case II}, when the coefficient of $P(x)$ in
(\ref{FOTP4}), i.e.
$[k_{+}f_{-}(x)+k_{-}f_{+}(x)]$,
has no zeroes in $[0,L)$,
one can solve (\ref{FOTP4}) for $P(x)$.
Note that Cases I and II do not necessarily exclude one another.\\
{\bf Case III}, when both the coefficients of $p(x)$ and
of $P(x)$ in Eq.
(\ref{FOTP4}) have zeroes,
{\em but not the same zeroes},
in $[0,L)$, and finally\\
{\bf Case IV}, when $f_{+}(x)$ and $f_{-}(x)$ have (at least) a common zero in
$[0,L)$. In this case one can directly notice that there is no net
flow of  particles, i.e., $J=0$, given the existence of the common
fixed point that cannot be crossed in any of the two
dynamics. This case will hence not be studied any further.

A detailed analysis reveals that the results obtained in
Cases I and II are completely equivalent and apply also
when $[f_{+}(x)-f_{-}(x)]$ and/or
$[k_{+}f_{-}(x)+k_{-}f_{+}(x)]$
have {\em different} zeroes in $[0,L]$,
i.e., in Case III. We therefore focus exclusively
on the main results pertaining to Case I.

In Case I
one can solve Eq.  (\ref{FOTP4}) with respect to $p(x)$, i.e.,
\begin{equation}
p(x)=\frac{J(k_{+}+k_{-})-
[k_{+}f_{-}(x)+k_{-}f_{+}(x)]P(x)}{f_{+}(x)-f_{-}(x)},
\label{case11}
\end{equation}
and by substitution in Eq.  (\ref{FOTP6}) one can obtain
a closed first-order differential equation  for $P(x)$:
\begin{eqnarray}
&&\left[f_{+}(x)f_{-}(x)\right]P'(x)+
\left\{\left[f_{+}(x)f_{-}(x)\right]'-
\left[f_{+}(x)f_{-}(x)\right]
\left[\mbox{ln}\left|f_{+}(x)-f_{-}(x)\right|\right]'+\right.\nonumber\\
&&+\left.[k_{+}f_{-}(x)+k_{-}f_{+}(x)]\right\}P(x)=
J\left\{(k_{+}+k_{-})+[f_{+}(x)-f_{-}(x)]
\displaystyle
\left[\frac{f_{+}(x)+f_{-}(x)}{2[f_{+}(x)-f_{-}(x)]}\right]^{'}
\right\}
.
\nonumber\\
\label{case12}
\end{eqnarray}
Recall that $(...)'$ denotes derivation with respect to $x$.

The crux of the problem resides in finding the
{\it correct} solution to (\ref{case12}) \cite{bronshtein}.
A blind application of 
the standard method of
variation of parameters leads to the familiar solution
\begin{equation}
P(x)=\displaystyle\frac{\langle\dot{x}\rangle}{L}\;\;
\displaystyle\left|\frac{f_{+}(x)-f_{-}(x)}{f_{+}(x)f_{-}(x)}
\right|
\left[CG(x,x_0)+K(x,x_0;x)\right],
\label{case13}
\end{equation}
where $C$ is a constant of integration that arises from the general solution to
the homogeneous part of Eq.  (\ref{case12}), the second contribution is the
particular solution of the full inhomogeneous equation , $x_0$ is an arbitrary
point in $[0,L)$, and we have defined the functions:
\begin{eqnarray}
G(u,v)&=&\exp\left\{-\int_v^udz\left[\displaystyle
\frac{k_{+}}{f_{+}(z)}+\frac{k_{-}}{f_{-}(z)}\right]\right\},
\nonumber\\
K(u,v;w)&=&\int_v^udz\;\mbox{sgn}
\left[\frac{f_{+}(z)f_{-}(z)}{f_{+}(z)-f_{-}(z)}\right]
\left[\displaystyle
\frac{k_{+}+k_{-}}{f_{+}(z)-f_{-}(z)}+
\left(\frac{f_{+}(z)+f_{-}(z)}{2\left[f_{+}(z)-f_{-}(z)\right]}\right)
'\right]G(w,z).
\nonumber
\label{case14}
\end{eqnarray}
The point is now to get the
{\it correct} integration constant $C$, or rather {\it constants}, since
one can not apply the above solution at the
points where the differential equation for $P(x)$
is singular, which is precisely at the location of fixed points,i.e., at
the zeroes of
$f_\pm$.


\subsection{No fixed points: $f_{+}(x)f_{-}(x)\neq 0$ in $[0,L)$}
\label{case1a}

In this case
there is no fixed point in any of the alternating dynamics ``+" and ``--".
The usual procedure to determine
$C$  in Eq.  (\ref{case13}) can now be followed, i.e., we require
periodicity of $P(x)$,
i.e.,
$P(x)=P(x+L)$, recalling
that both $f_{+}(x)$ and $f_{-}(x)$ are periodic.
This results in the expression Eq.  (\ref{simP}) of the
main text.
The mean velocity
at the steady state, cf. Eq. (\ref{simvelocity}) on
the main text, follows from the normalization of $P(x)$.


\subsection{One of the alternating dynamics has two fixed points in $[0,L)$}
\label{case1b}

The situation is entirely different, both physically and mathematically,
when the
system can cross {\em unstable} fixed points within the interval $[0,L)$
in the
long time limit. Consider the simple case described in Section \ref{mainb} in
the main text. Clearly, the steady state results leading to Eqs.  (\ref{FOTP4})
and (\ref{FOTP6}) still apply, but the solution of Eq.  (\ref{FOTP6}) is more
delicate than the ``blind" integration that
led to expression
(\ref{case13})
for $P(x)$. Indeed, in the vicinity of the fixed points
$x_1$ and $x_2$,
if we simply try to apply formula (\ref{case13}), we are led to the
dependence
\begin{equation}
P(x)\sim
{|x-x_{1,2}|^{-1-k_{+}/[f'_{+}(x_{1,2})]}}.
\label{case19}
\end{equation}
For the case of the stable fixed point, $P(x)$ is therefore
continuous when $k_{+}/|f'_{+}(x_1)| > 1$, and divergent but integrable
for  $k_{+}/|f'_{+}(x_1)| \leqslant 1$, a result that causes no conceptual
difficulties. However, at the
unstable fixed point $x_2$, this procedure leads to an apparent non-integrable
divergence, which is clearly unphysical, and mathematically improper in
view of the
requirement of normalization of $P(x)$.

As explained in detail in \cite{shortPRE},
the fallacy lies in the assumption that a single integration constant $C$,
see Eq.  (\ref{case13}), is valid throughout the region $[0,L]$. One solves
the problem by choosing different integration constants
in each of the separate
intervals $[0,x_1)$, $(x_1,x_2)$, and $(x_2,L]$ between the fixed
points.
There is {\em exactly one} choice of this constant valid for both
$(x_1,x_2)$ and
$(x_2,L)$ such that the divergence at $x_2$ is removed, namely
$C=-K(x_2,x_0; x_0)$;
and another choice valid in the
interval $[0,x_1)$ that ensures the required continuity
and periodicity of $P(x)$.
The acceptable expression for the probability density is
therefore found to be:
\begin{equation}
P(x)=
\left\{
\begin{array}{ll}
\displaystyle\frac{\langle\dot{x}\rangle}{L}\;\;
\displaystyle\left|\frac{f_{+}(x)-f_{-}(x)}{f_{+}(x)f_{-}(x)}\right|
\left[K(L,x_2;L)G(x,0)+K(x,0;x_0)\right]
&\mbox{for}\,\,\,\,x \in [0,x_1)\\
\\
\displaystyle\frac{\langle\dot{x}\rangle}{L}\;\;
\displaystyle\left|\frac{f_{+}(x)-f_{-}(x)}{f_{+}(x)f_{-}(x)}\right|
K(x,x_2;x),&\mbox{for}\,\,\,\,x \in (x_1,L).
\end{array}
\right.
\label{case110}
\end{equation}

These expressions can be further simplified if one takes as the basic
period not
$[0,L]$, but $[x_1,x_1+L]$. Then the simple-looking, ``compact"
expression
$P(x)=\displaystyle\frac{\langle\dot{x}\rangle}{L}\;\;\displaystyle
\left|\frac{f_{+}(x)-f_{-}(x)}{f_{+}(x)f_{-}(x)}\right|
K(x,x_2;x)$ [Eq.  (\ref{twopointsP}) in the main text] holds throughout
this new basic period.
With this choice
of the basic interval, the normalization condition
$\displaystyle\int_{x_1}^{x_1+L} P(x)dx=1$ leads to the expression
(\ref{case116}) for the mean velocity.

$P(x)$ as given above [Eq. (\ref{case110}) or Eq. (\ref{twopointsP})] meets all
the requirement enumerated in Sec. \ref{FOTP}. In particular, let us check its
behavior at the fixed points $x_1$ and $x_2$. 
In order to do this, it is useful
to write Eq. (\ref{twopointsP}) as:
\begin{equation}
P(x)=\displaystyle\frac{\langle\dot{x}\rangle}{L}\;\;\displaystyle
\left|\frac{f_{+}(x)-f_{-}(x)}{f_{-}(x)}\right|\,
\displaystyle\frac{K(x,x_2;0)}{\left|f_{+}(x)\right|\,G(0,x)}\,.
\end{equation}
For $x=x_2$ (the unstable fixed point), 
${K(x,x_2;0)}/\left\{\left|f_{+}(x)\right|\,G(0,x)\right\}$ 
presents an indeterminacy of the type ``$0/0$"; applying H\^{o}spital's rule
one simply finds that $P(x)$ is continuous in $x=x_2$, and its value is given by
Eq.~(\ref{twopointsx2}).

For $x=x_1$ (the stable fixed point), there are three situations that might
appear, depending on the value of $k_+/\left|f_+'(x_1)\right|$:\\
(a) For $k_+/\left|f_+'(x_1)\right|>1$, both ${K(x,x_2;0)}$ and 
$\left|f_{+}(x)\right|\,G(0,x)$ present a divergence for $x\searrow x_1$,
$x\nearrow x_1+L$; therefore, $P(x)$ presents an indeterminacy of the type 
``$\infty/\infty$". Applying H\^{o}spital's rule one finds that $P(x)$
has the same finite limit as $x\searrow x_1$,
$x\nearrow x_1+L$, as indicated in Eq.~(\ref{case113}).\\
(b) For $k_+/\left|f_+'(x_1)\right|<1$, ${K(x,x_2;0)}$ is finite
and nonzero at $x=x_1,x_1+L$, while $\left|f_{+}(x)\right|\,G(0,x)
\rightarrow 0$ as $\left|x-x_1\right|^{1-k_+/\left|f_+'(x_1)\right|}$,
(respectively $\left|x-x_1-L\right|^{1-k_+/\left|f_+'(x_1)\right|}$). 
Therefore, $P(x)$ presents a power-law integrable divergence, 
$P(x)\sim \left|x-x_1\right|^{-1+k_+/\left|f_+'(x_1)\right|}$ in the neighborhood of $x_1$,
respectively $P(x)\sim \left|x-x_1-L\right|^{-1+k_+/\left|f_+'(x_1)\right|}$
in the neighborhood of $x_1+L$.\\
(c) Finally, for $k_+/\left|f_+'(x_1)\right|=1$, ${K(x,x_2;0)}$ behaves like 
$\mbox{ln}\left|x-x_1\right|$ in the vicinity of $x_1$
(respectively, like $\mbox{ln}\left|x-x_1-L\right|$ in the vicinity of $x_1+L$),
while $\left|f_{+}(x)\right|\,G(0,x)$ has a finite limit
in these points. Thus, $P(x)$ has a
``marginal", logarithmic-like divergence in these points.

\subsection{Each of the two alternating dynamics has two fixed points in
$[0,L)$}
\label{case1c}

Suppose that each of the two dynamics, ``+" and ``--", has two fixed
points
(one stable and one unstable)
in the interval $[0,L)$, i.e., the situation described in
Section \ref{mainb}.  Again, if one blindly applies 
the result (\ref{case13}) for $P(x)$, then
in the vicinity of the fixed points
\begin{eqnarray}
P(x-x_{1,2})&\sim&
|x-x_{1,2}|^{-1-k_{+}/[f'_{+}(x_{1,2})]},
\nonumber\\
P(x-x_{3,4})&\sim&
|x-x_{3,4}|^{-1-k_{-}/[f'_{-}(x_{3,4})]}.
\label{case119}
\end{eqnarray}
One again encounters 
the nonphysical non-integrable divergence at the {\em
unstable} fixed points.
The correct procedure is again to use the  solution
(\ref{case13}), but with  different integration constants $C$ in each
of the
open intervals between the fixed points. The latter
constants are determined by the requirements  (i) to remove the strong
divergences at
the unstable fixed points, by imposing the condition that the coefficient
of the divergent term becomes zero
at these points; (ii) to ensure continuity and periodicity of $P(x)$;
and finally, through the normalization condition, (iii) to determine
the flow
$J$ and the mean asymptotic velocity $\langle\dot{x}\rangle$.

Case a: $0\leqslant x_1$ (s) $<x_3$ (u) $<x_4$ (s) $<x_2$ (u) $<L$\\
The above ``program" leads to the expression (\ref{case120}) in the
main text for the
probability density (in the appropriately chosen basic period
$[x_1,x_1+L]$), and to the corresponding
mean asymptotic velocity, Eq.  (\ref{case124}).

Case b: $0\leqslant x_1$ (s) $<x_2$ (u) $<x_3$ (s) $<x_4$
(u) $<L$\\
One gets the
following expression for the
probability density, again in the basic interval $[x_1,x_1+L]$:
\begin{equation}
P(x)=
\left\{
\begin{array}{l}
\displaystyle\frac{\langle\dot{x}\rangle}{L}\;\;
\displaystyle\left|\frac{f_{+}(x)-f_{-}(x)}{f_{+}(x)f_{-}(x)}\right|
K(x,x_2,x),
\mbox{for}\,\,\,\,x\in(x_1,x_3) \\
\\
\displaystyle\frac{\langle\dot{x}\rangle}{L}\;\;
\displaystyle\left|\frac{f_{+}(x)-f_{-}(x)}{f_{+}(x)f_{-}(x)}\right|
K(x,x_4,x),
\mbox{for}\,\,\,\,x\in(x_3,x_1+L)
\end{array}
\right.,
\label{case125}
\end{equation}
with the good behaviors (continuity) at the unstable fixed points $x_2$ and
$x_4$
and either continuity, or (integrable) divergences at the stable fixed points
$x_1$ and $x_3$; as well as the required periodicity for $P(x)$.
The mean asymptotic velocity is therefore given by:
\begin{equation}
\langle\dot{x}\rangle=L\left\{\int_{x_1}^{x_3}dx\displaystyle
\left|\frac{f_{+}(x)-f_{-}(x)}{f_{+}(x)f_{-}(x)}\right|K(x,x_2,x)
+
\int_{x_3}^{x_1+L}dx\displaystyle
\left|\frac{f_{+}(x)-f_{-}(x)}{f_{+}(x)f_{-}(x)}\right|
K(x,x_4,x)
\right\}^{-1}.
\end{equation}

\acknowledgments{We acknowledge support from the Swiss National Science
Foundation (I.B.), the National Science Foundation under Grant
Nos. PHY-9970699 (K.L.)
and DMS-0079478 (R.K.) and the STOCHDYN program of the European Science
Foundation (C.V.d.B.). We also thank Profs. P. Talkner and P. H\"anggi
for drawing our attention to Refs. \cite{behn,zapata,bronshtein}.}

\end{document}